\documentclass{webofc}
\usepackage[varg]{txfonts}   
\begin{document}
\title{Sound velocity, equation of state and strangeness \\in neutron star matter}
%
%

\author{\firstname{Wolfram} \lastname{Weise}\inst{1}\fnsep\thanks{\email{weise@tum.de}}
}

\institute{Technical University of Munich,  TUM School of Natural Sciences, Physics Department, \\85747 Garching, Germany
          }

\abstract{%
This presentation in two parts starts with a brief review on the speed of sound in neutron star matter as inferred from observational data.  It is pointed out that, in view the strong constraints imposed by observed properties of the heaviest neutron stars,  the equation of state must be very stiff.  Stringent limits can be set on the possible appearance of phase transitions in neutron star cores.  The second part discusses issues concerning the occurance of strangeness degrees of freedom,  in particular of hyperons,  in neutron stars.  Here a focus is on the role of repulsive hyperon-nuclear three-body forces,  potentially resolving the so-called hyperon puzzle.
}
\maketitle
\section{Introduction}
\label{intro}

Within the last decade,  the observational data base for neutron stars has expanded substantially.  Its detailed analysis sharpened the empirical constraints on the equation-of-state (EoS) of the matter deep inside the core of such extreme objects.  In particular,  the discovery of two-solar-mass neutron stars implied that the EoS,  i.e.  pressure $P(\varepsilon)$ as function of energy density $\varepsilon$,  must be sufficiently stiff in order to support such heavy compact stars against gravitational collapse.  Some previously discussed simple forms of exotic matter (quark matter,  kaon condensation, ...) were thus excluded when their corresponding EoS's turned out to be too soft and unable to satisfy the stability conditions.

It has long been considered that the formation of hyperons in the cores of neutron stars through weak processes might become energetically favourable at sufficiently high densities.  However, it was soon realized that when adding these strangeness degrees of freedom,  the EoS would again become too soft and unable to support $2 M_\odot$ neutron stars.  This started a still continuing discussion under the keyword {\it hyperon puzzle}. To frame this discussion in a broader context,  it is useful first to summarize the present state of knowledge about the neutron star EoS as inferred directly from the observational data.  Of particular interest is the speed of sound in the core of neutron stars.  Its behaviour as a function of energy density is a sensitive indicator for phase transitions or continuous changes of degrees of freedom (crossovers) in the star's composition. 

\section{Neutron star matter equation of state}
\label{sec-2}
\subsection{Observational constraints}
\label{sec-2.1}

{\it Neutron star masses and radii}.  The information about the masses of the heaviest neutron stars derives primarily from precise Shapiro time delay measurements of pulsars orbiting in binary systems with white dwarfs.  Three such massive objects, PSR J1614–2230 \cite{Demorest2010,Fonseca2016,Arzoumanian2018}, PSR J0348+0432 \cite{Antoniadis2013} and PSR J0740+6620 \cite{Cromartie2020,Fonseca2021},  have been established in the past:
\begin{eqnarray}
	&\text{PSR J1614–2230} \qquad &M = 1.908 \pm 0.016 \, M_\odot ~, ~ \label{eq:ShapiroMass1}\\
	&\text{PSR J0348+0432} \qquad &M = 2.01 \pm 0.04 \, M_\odot ~, \label{eq:ShapiroMass2}\\
	&\text{PSR J0740+6620} \qquad &M = 2.08 \pm 0.07 \, M_\odot ~. \label{eq:ShapiroMass3}
\end{eqnarray}
Recently the heaviest neutron star observed so far has been reported \cite{Romani2022}:
\begin{eqnarray}
	&\text{PSR J0952-0607} \qquad &M = 2.35\pm 0.17 \, M_\odot ~.~ \label{eq:BWMass}
	\end{eqnarray}
This is also one of the fastest rotating pulsars.  With a spin period of 1.4 ms it requires corrections for rotational effects as described in \cite{Brandes2023a}. 

Radii of neutron stars together with their masses can be inferred from X-ray profiles of rotating hot spot patterns measured with the NICER telescope,  also in combination with other multimessenger data.  Two representative neutron stars have been investigated in this way \cite{Riley2019,Riley2021}: 
\begin{eqnarray}
	\text{PSR J0030+0451} &\qquad M = 1.34^{+0.15}_{-0.16} \, M_\odot ~, \qquad &R=12.71^{+1.14}_{-1.19}\,\text{km}~,\label{eq:NICER1}\\
         \text{PSR J0740+6620} &\qquad M = 2.072^{+0.067}_{-0.066} \, M_\odot ~, \qquad &R=12.39^{+1.30}_{-0.98}\,\text{km}~.\label{eq:NICER1}
\end{eqnarray}
These combined mass and radius data are included as prime input in setting constraints for the neutron star EoS.\\
\\
{\it Binary neutron star mergers and tidal deformabillities}.  Gravitational wave signals produced by the merger of two neutron stars in a binary have been detected by the LIGO and Virgo collaborations \cite{Abbott2019}.  These signals are interpreted using theoretical waveform models that depend on the mass ratio of the two neutron stars, $M_2/M_1$,  and a combination of their tidal deformabilities.  Based on information from the GW170817 event a (dimensionless) tidal deformability $\Lambda_{1.4} = 190^{+390}_{-120}$ has been deduced for a 1.4 solar mass neutron star \cite{Abbott2018}.  

\subsection{Inference of sound velocity and EoS in neutron stars}
\label{sec-2.2}

The data listed in the previous subsection have been incorporated in a variety of studies using Bayesian statistical methods to infer the squared sound velocity, $c_s^2(\varepsilon) = \partial P/\partial\varepsilon$,  and the equation of state,  $P(\varepsilon)= \int_0^\varepsilon d\varepsilon'\,c_s^2(\varepsilon')$, of neutron star matter and related properties \cite{Brandes2023, Brandes2023a}.  Constraints at low baryon densities are commonly imposed by implementing results from nuclear chiral effective field theory (ChEFT).  In our recent work \cite{Brandes2023,Brandes2023a} this constraint is used in the Bayes inference procedure as a likelihood ({\it not} as a prior) within a conservative window of baryon densities,  $\rho \lesssim 1.3\,\rho_0$,  around the equilibrium density $\rho_0 = 0.16$ fm$^{-3}$ of normal nuclear matter.  At asymptotically high densities the conformal limit,  $c_s^2\rightarrow 1/3$,  is implemeted as given by perturbative QCD. 

\begin{figure}
\centering
\sidecaption
\includegraphics[width=7cm,clip]{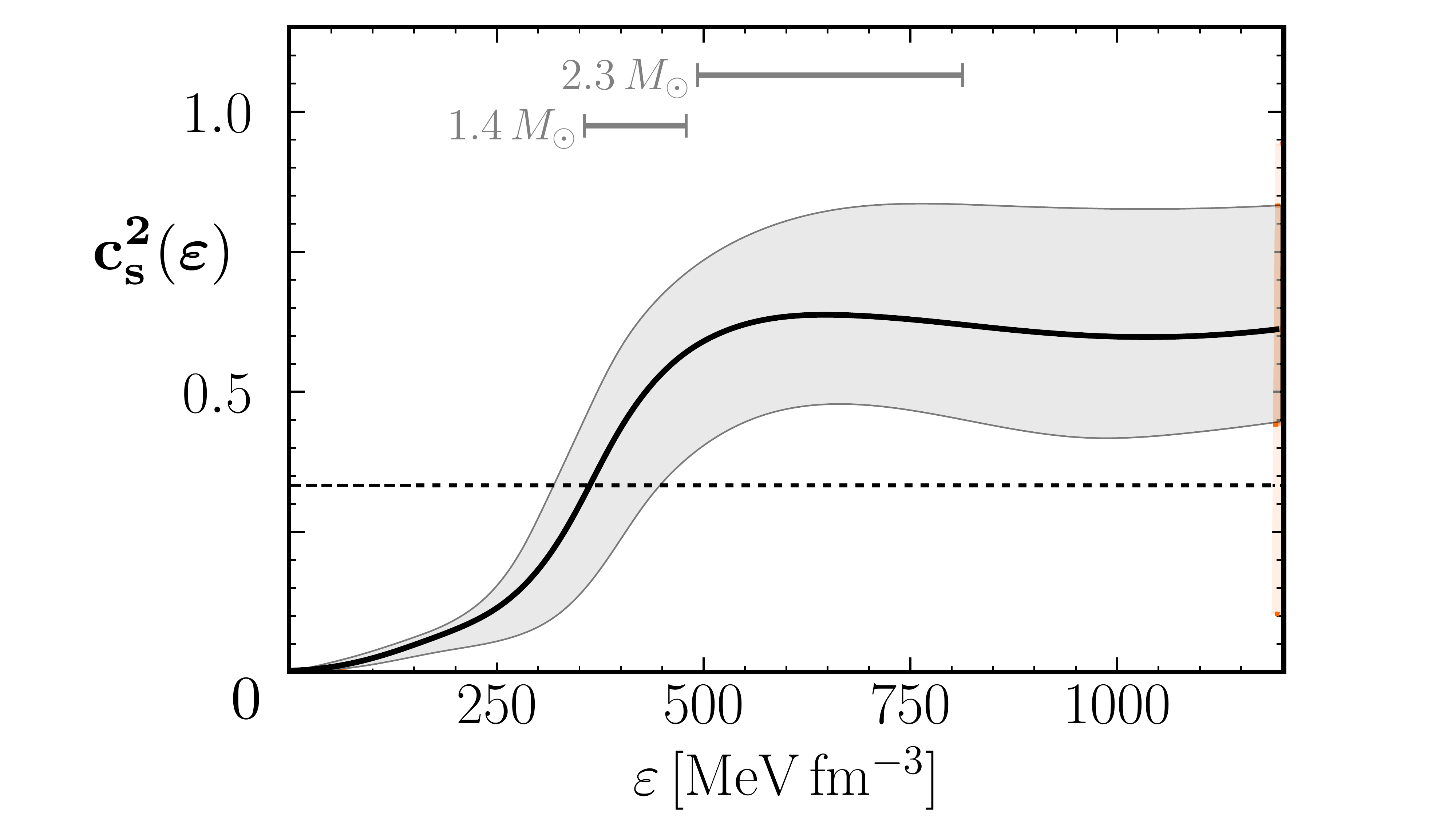}
\caption{Squared speed of sound \cite{Brandes2023a} as function of energy density inferred from the empirical data set listed in Section\,\ref{sec-2.1}.  Median (solid curve) and posterior probability distribution at 68\% confidence level (grey band) are displayed.  Intervals are indicated for ranges of central energy densities in the cores of 1.4 and 2.3 $M_\odot$ neutron stars.}
\label{fig-1}       
\end{figure}

The result for the inference of the sound velocity in Fig.\,\ref{fig-1} shows a rapid increase of $c_s^2$ beyond the conformal bound of 1/3 in the range of energy densities relevant for a wide range of neutron stars with masses $M \sim$ 1.4 - 2.3 $M_\odot$.  Fig.\,2 presents the inferred EoS band.  Notably,  the inclusion of the new heavy 2.3 $M_\odot$ black widow pulsar PSR J0952-0607 in the data set leads to a rapidly rising pressure,  even exceeding that of the time-honored APR EoS \cite{APR1998}.

\begin{figure}
\centering
\sidecaption
\includegraphics[width=7.2cm,clip]{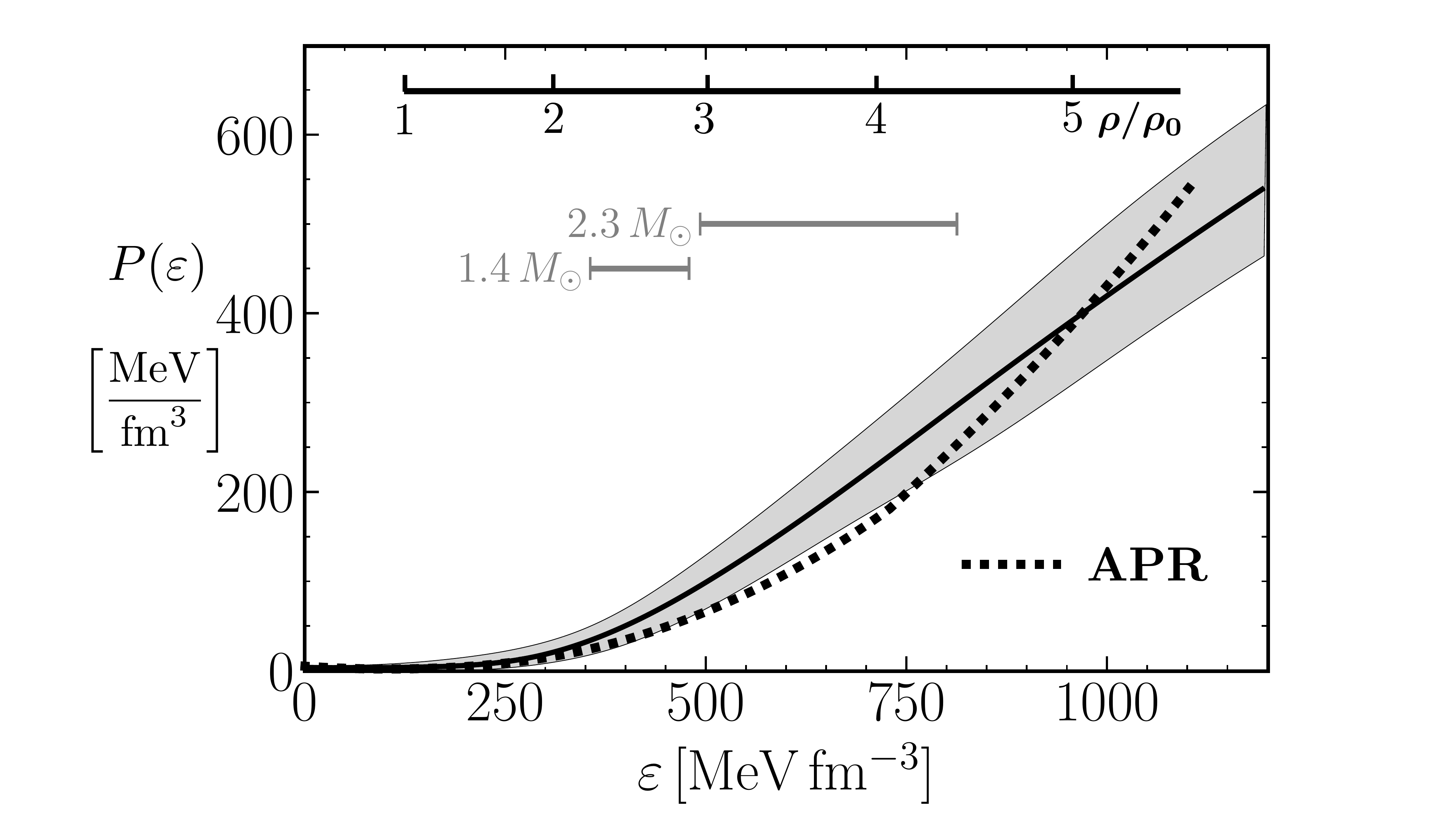}
\caption{Equation of state $P(\varepsilon)$ deduced \cite{Brandes2023a} from the inferred sound velocity based on the empirical data set listed in Section \,\ref{sec-2.1}. Median and 68\% confidence band as in Fig.\,\ref{fig-1}.  Also shown for orientation is the baryon density, scale, $\rho/\rho_0$,  in units of the equilibrium density of nuclear matter, $\rho_0 = 0.16$ fm$^{-3}$,  and the APR EoS \cite{APR1998} (dotted line) for comparison.}
\label{fig-2}       
\end{figure}

\begin{figure}
\centering
\sidecaption
\includegraphics[width=7cm,clip]{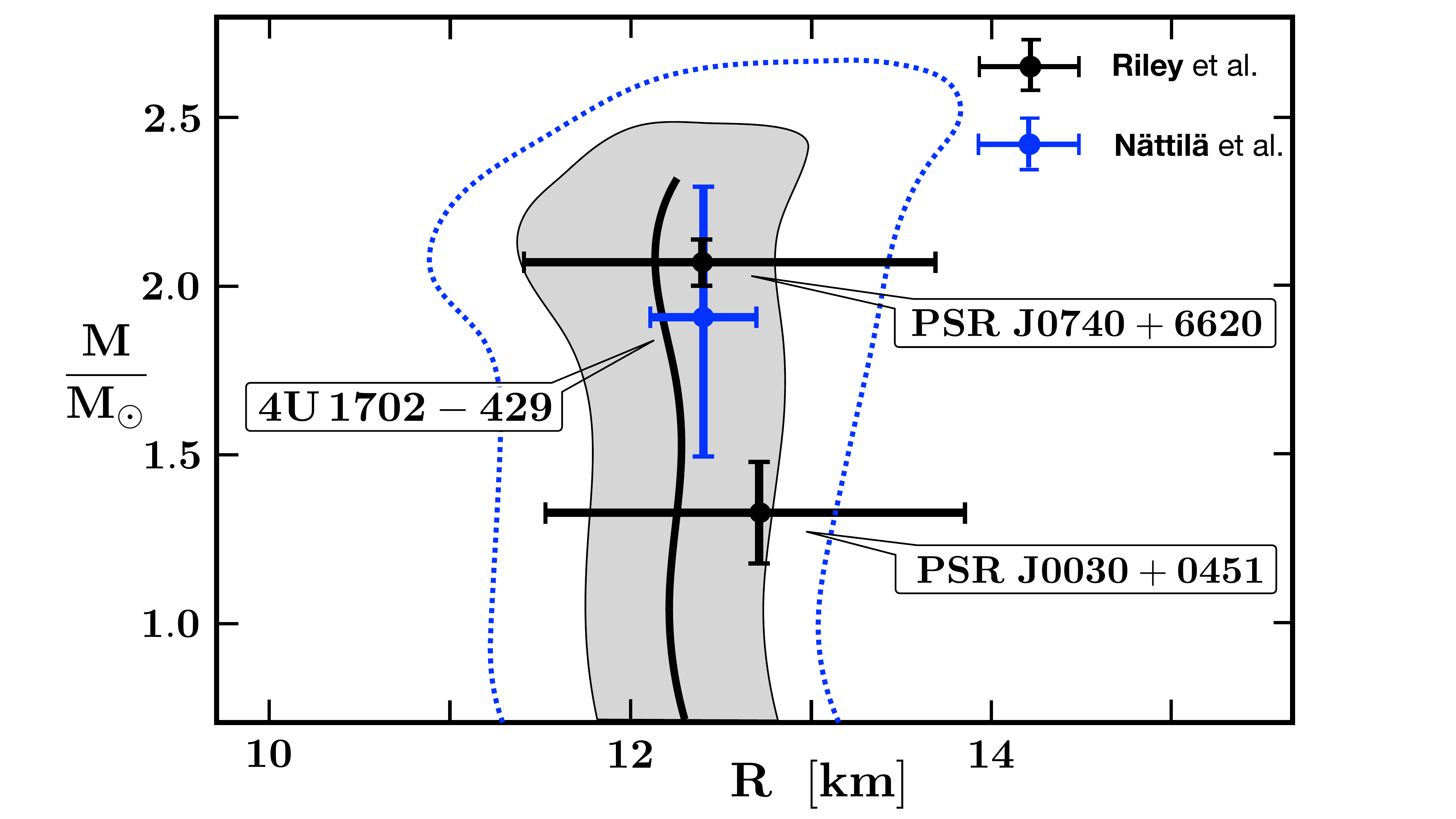}
\caption{Posterior probability distribution \cite{Brandes2023a} of the mass-radius relation at 68\% level (grey band) and 95\% level (dotted border line), compared to the NICER data analysis by Riley {\it et al.} of PSR J0030+0451 and PSR J0740+6620 \cite{Riley2019, Riley2021}.  In addition the 68\% mass-radius credible interval of the thermonuclear burster 4U 1702-429 is displayed \cite{Nattila2017} (blue).}
\label{fig-3}      
\end{figure}

The Bayes inference computations involve about $10^6$ EoS samples.  Their confrontation with data through solution of the Tolman-Oppenheimer-Volkov equations generates posterior bands for the mass-radius relation of neutron stars as displayed in Fig.\,\ref{fig-3}.  The  profiles computed for neutron stars in the mass range 1.4 - 2.3 $M_\odot$ reach baryon densities of always less than $5\,\rho_0$ in the central cores of even the heaviest stars.  For example,  $\rho_c(1.4\,M_\odot) = (2.6\pm 0.4)\,\rho_0$ and $\rho_c(2.3\,M_\odot) = (3.8\pm 0.8)\,\rho_0$ at 68\% confidence level \cite{Brandes2023a}.  In a baryonic picture of the neutron star,  this implies that the average distance between baryons even in the highly compressed center always exceeds 1 fm,  more than twice the characteristic hard-core distance of 1/2 fm in the nucleon-nucleon interaction.

The relatively moderate baryon densities together with a Bayes factor analysis for the likelihood of very small sound speeds and limiting constraints for the maximally possible extension of Maxwell-constructed phase coexistence regions suggest that the appearance of a strong first-order phase transition in neutron star cores is extremely unlikely \cite{Brandes2023a}.  On the other hand,  a hadron-quark continuity crossover \cite{Baym2019, McLerran2019, Fukushima2020} is not excluded.

\section{Strangeness and baryonic matter}
\label{sec-3}
\subsection{Hyperon-nucleon interactions from SU(3) chiral effective field theory}
\label{sec-3.1}

Guided by the symmetry breaking pattern of three-flavor QCD,  chiral SU(3) effective field theory (ChEFT) is the method of choice to investigate interactions involving the baryon and pseudoscalar meson octets.  Our present (still limited) quantitative understanding of hyperon-nucleon interactions derives primarily from this framework at next-to-leading order (NLO) \cite{Haidenbauer2013, Haidenbauer2020, Petschauer2020}.  The NLO19 version of the interaction  \cite{Haidenbauer2020} has been checked successfully against an increasing amount of data.  It now serves as a prototype force for $\Lambda N$,  $\Sigma N$ and $\Xi N$ two-body processes,  including important coupled-channel dynamics such as $\Lambda N \leftrightarrow \Sigma N$ and $\Xi N\leftrightarrow \Lambda\Lambda$. Steps towards NNLO have recently been taken \cite{Haidenbauer2023}. 

{\it Hyperon-nucleon scattering and correlation functions}.  Hyperon-nucleon cross sections measured decades ago at BNL and CERN served as a first step in constraining the theory,  more recently strengthened by data from J-PARC and JLab.  Further input comes from the femtoscopy studies of hyperon-nucleon correlation functions from $pp$ collisions with ALICE at LHC \cite{Fabbietti2021}.  In particular,  the accurately determined $\Lambda p$ correlation function \cite{Acharya2021} is well reproduced by the NLO19 interaction,  especially at the lowest center-of-mass momenta ($\lesssim 100$ MeV/c) which are not covered by scattering experiments. 

{\it Constraints from hypernuclei}.  Hypernuclear physics has provided a systematic set of binding energies for ground states and excited states of a $\Lambda$ in nuclei from A = 5 to 208.  A phenomenological $\Lambda$-nuclear single-particle potential works well to reproduce the data with a central potential depth $U_\Lambda(r=0) \simeq -30$ MeV \cite{Gal2016}.  Hypernuclear many-body calculations,  using e.g. Brueckner-Hartree-Fock methods \cite{Haidenbauer2020b, Gerstung2020},  have henceforth addressed the question to what extent such a shell-model potential can be understood in terms of underlying $\Lambda N$ two-body forces.  The results of those computations point to a trend, namely that a realistic two-body interaction produces too much binding in the lower shell-model levels of heavy hypernuclei such as $_\Lambda Pb$ \cite{Haidenbauer2020b}.  Similarly,  the $\Lambda$ potential in nuclear matter calculated with the NLO19 two-body force alone is more attractive by almost 20\% than the empirical potential and has its minimum located at a density 25\% higher than $\rho_0 \simeq 0.16$ fm$^{-3}$ \cite{Gerstung2020}. 

This opens the question about the role of $\Lambda NN$ three-body forces. 
The issue is further underlined by a recent phenomenological study \cite{Friedman2023} using an ansatz that combines contributions of $\Lambda$-nucleon two- and three-body forces to the $\Lambda$-nuclear potential: $U_\Lambda(\rho) = U_0^{(2)} [\rho(r)/\rho_0 ]+ U_0^{(3)} [\rho(r)/\rho_0]^2$.  An optimal fit to the systematics of hypernuclear binding energies of the lowest ($s$ and $p$ shell) orbits has been performed. The best-fit result features a total potential depth of $U_\Lambda(\rho = \rho_0) = -(27.3\pm 1.6)$ MeV,  separated into a $\Lambda N$ two-body contribution with inclusion of important Pauli corrections,  $U_0^{(2)} = -(38.6\pm 0.8)$ MeV,  and a three-body piece,  $U_0^{(3)} = (11.3\pm 1.4)$ MeV,  suggesting the presence of a substantial repulsive $\Lambda NN$ three-body force.

{\it Hyperon-nuclear three-body forces}. Within the ChEFT hierarchy,  genuine three-baryon interactions first appear at next-to-next-to-leading order (NNLO) \cite{Petschauer2016}. A complete calculation would involve a large number of parameters, too large for a reliable determination given the limited amount of data.  However,  an approximation known to work well in the three-nucleon sector,  namely $\Delta(1232)$ dominance in intermediate states of the three-body mechanism,  can easily be generalized to decuplet dominance in SU(3) ChEFT,  now with $\Sigma^*$ and $\Delta$ intermediate virtual excitations.  In this way the three-body terms are promoted from NNLO to NLO in the chiral counting,  in line with the NLO treatment of the two-body forces.  This leaves three coupling parameters of the three-body interaction terms.  Assuming SU(3) symmetry,  one of these coupling constants is already fixed by the $\Delta \rightarrow \pi N$ decay width.  The remaining two parameters can be constrained by comparison with the empirical $\Lambda$-nuclear single-particle potential \cite{Gerstung2020}.

\subsection{Hyperons in dense baryonic matter and neutron stars}
\label{sec-3.2}

Equiped with a $\Lambda NN$ three-body force from chiral SU(3) ChEFT,  one can now proceed to investigate its role in nuclear and neutron star matter.  

\begin{figure}
\centering
\includegraphics[width=11.0cm,clip]{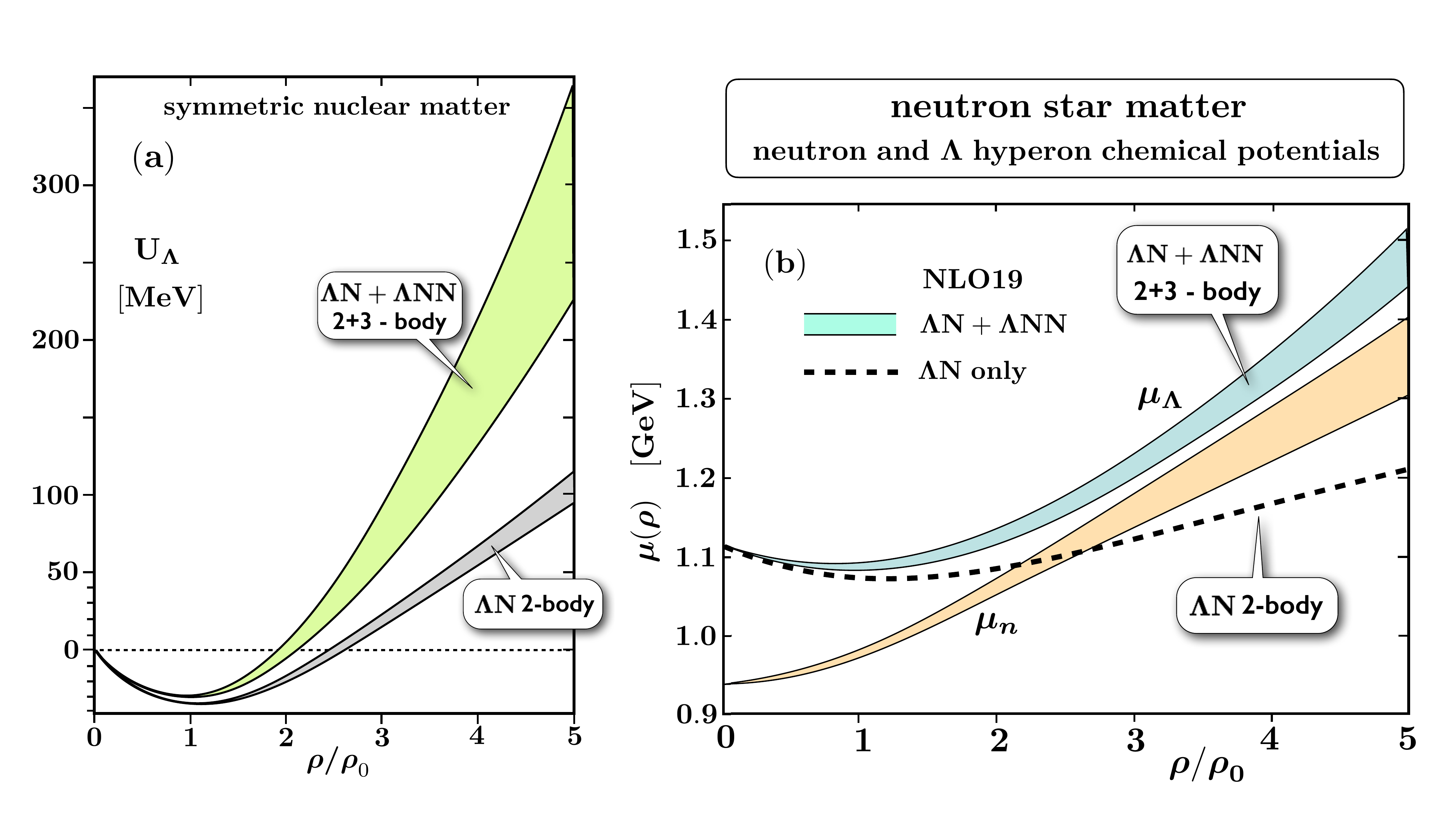}
\caption{(a) Single-particle potential $U_\Lambda$ of a $\Lambda$ hyperon in symmetric nuclear matter as a function of baryon density in units of $\rho_0 = 0.16$ fm$^{-3}$.  Results of Brueckner-Hartree-Fock calculations using NLO19 two-body $\Lambda N$ interaction, in comparison with additional $\Lambda NN$ three-body forces as input.  Three-body parameters have been fixed to the empirical $U_\Lambda(\rho_0) \simeq -30$ MeV.  (b) Chemical potentials of $\Lambda$ and neutron, $\mu_\Lambda$ and $\mu_n$ in beta-equlibrated neutron star matter calculated with the same input as in (a).  The neutron chemical potential is computed using the chiral nucleon-meson model combined with functional renormalization group methods \cite{Drews2017}.  In both Figures the bands give an impression of estimated uncertainties.  Adapted from \cite{Gerstung2020}.}
\label{fig-4}   
\end{figure}

{\it Density dependence of the $\Lambda$-nuclear potential}.  With increasing baryon density beyond $\rho\gtrsim\rho_0$ the repulsive nature of the $\Lambda NN$ three-body interaction is expected to become progressively more prominent relative to the contribution of the $\Lambda N$ two-body interaction,  because it enters the energy density with a higher power of density.  This is shown in Figure\,\ref{fig-4}(a) for the $\Lambda$-nuclear single-particle potential,  $U_\Lambda(\rho)$, in symmetric nuclear matter.  A smilar behaviour is observed for a $\Lambda$ hyperon in neutron matter.  At high densities,  $\rho \gtrsim 3\,\rho_0$,  the three-body contribution to $U_\Lambda$ begins to be dominant over the two-body part with its linear density dependence characteristic of a Hartree potential generated by the short-distance $\Lambda N$ interaction.

{\it Hyperon puzzle in neutron stars}.  Given the energy density $\varepsilon(\rho_i)$ of neutron star matter as a function of the densities $\rho_i$ of baryonic constituents (neutrons, protons and possibly $\Lambda$ hyperons),  the individual chemical potentials are determined by $\mu_i = \partial\varepsilon / \partial\rho_i$.  The onset for the appearance of hyperons in a neutron star through weak interaction processes converting neutrons into $\Lambda$'s is marked by the condition that the $\Lambda$ chemical potential equals that of the neutrons: $\mu_\Lambda = \mu_n$.

The density dependence of the potential $U_\Lambda$ is reflected in $\mu_\Lambda$.  Hence the repulsive $\Lambda NN$ three-body interaction shows up prominently in the $\Lambda$ chemical potential at high baryon densities.  Figur\,\ref{fig-4}(b) displays results of a Brueckner-Hartree-Fock calculation of $\mu_\Lambda$ for neutron star matter in comparison to $\mu_n$ \cite{Gerstung2020}.  If only two-body $\Lambda N$ interactions (NLO19) are included,  hyperons start appearing already at densities $\rho \simeq 2-3\,\rho_0$. The EoS becomes too soft and unable to support two-solar-mass neutron stars.  However,  repulsive $\Lambda NN$ three-body forces raise $\mu_\Lambda$ to a level such that it does not match the neutron chemical potential any more: the appearance of $\Lambda$'s in the core of the neutron star is prohibited.  The ChEFT-based $\Lambda NN$ three-body potential follows closely the expected ansatz $U^{(3)}(\rho) = U_0^{(3)} (\rho/\rho_0)^2$.  One finds $U_0^{(3)}\simeq 10$ MeV,  consistent with the result of systematic hypernuclear phenomenology \cite{Friedman2023}.  At the same time the neutron sector (with inclusion of many-body correlations) features sufficiently strong repulsion by itself to satisfy the stabilty conditions for heavy neutron stars.  Hence while the final solution of the {\it hyperon puzzle} continues to be under discussion,  the results so far point into an at least qualitatively promising direction.\\

{\it Acknowledgement:} Work supported in part by the DFG Excellence Cluster ORIGINS.

\end{document}